# Stochastic Boltzmann Equation for Magnetic Relaxation in High-Spin Molecules

Daniel M. Packwood[1], Helmut G. Katzgraber[2,3,4,5], and Winfried Teizer[2,1,3]

1 Advanced Institute for Materials Research, Tohoku University, 2-1-1 Katahira, Aoba-ku, Sendai 980-8577, Japan

2 Department of Physics and Astronomy, Texas A & M University, College Station, TX 77843-4242, USA

3 Materials Science and Engineering Program, Texas A & M University, College Station, TX 77843, USA

4 Santa Fe Institute, 1399 Hyde Park Road, Santa Fe, NM 87501, USA

5 Applied Mathematics Research Centre, Coventry University, Coventry, CV1 5FB, England

**Abstract**

We introduce the stochastic Boltzmann equation (SBE) as an approach for exploring the spin dynamics of magnetic molecules coupled to a stochastic environment. The SBE is a time-evolution equation for the probability density of the spin density matrix of the system. This probability density is relevant to experiments which take measurements on single molecules, in which probabilities of observing particular spin states (rather than ensemble averages) are of interest. By analogy with standard treatments of the regular Boltzmann equation, we propose a relaxation-time approximation for the SBE, and show that solutions to the SBE under the relaxation-time approximation can be obtained by performing simple trajectory simulations for the case of a boson gas environment. Cases where the relaxation-time approximation are satisfied can therefore be investigated by careful choice of the parameters for the boson gas environment, even if the actual environment is quite different from a boson gas. The application of the SBE approach is demonstrated through an illustrative example.

## 1. Introduction

Magnetic relaxation of high-spin molecules coupled to an external environment is a topic of fundamental and practical importance[1]. Technological breakthroughs such as ultra-high density memory or even quantum computation are widely expected if high-spin molecules could be deposited onto conducting surfaces with their magnetic properties in-tact[2, 3, 4]. However, interactions between the molecule and the phonon modes of the surface are inevitable, leading to rapid magnetic relaxation of the molecule. Therefore, information stored by manipulating the direction of the molecule's spin vector is quickly lost[5, 6, 7, 8, 9, 10]. This leads to the important experimental goal of identifying conditions in which the rate of magnetic relaxation of the adsorbed molecule is relatively small. In turn, this encourages the development of new theoretical techniques that yield insights into the physical nature of the magnetic relaxation phenomenon in high-spin molecules[11, 12, 13, 14, 15, 16].

Experimental techniques such as high-resolution scanning tunneling microscopy have developed to the point where the spin states of individual molecules adsorbed to a surface can be probed[3, 17]. This is a significant advance over techniques where ensemble averages of bulk or film samples where measured[18, 19]. To see how this provides an interesting opportunity for theoretical research, consider the diagram shown in Figure 1a, which loosely depicts the low-energy ($S = 10$) spin states in the molecule $Mn_{12}$-ac (= $Mn_{12}O_{12}(CH_3COO)_{16}$)[20, 21]. These spin states are arranged in a double-well structure, which often occurs in real molecules due to the presence of an internal easy axis[1]. Tunneling between states in different wells may occur in the presence of external magnetic fields, and transitions within each well occur when the molecule is coupled to the degrees of freedom of an environment (e.g., the phonon modes of a surface)[22, 23]. Spin-relaxation can therefore be regarded as the net outcome of spin-tunneling between states in different wells, as well as transitions within each well. The key point is that the molecule can utilize a great variety of pathways when relaxing to its equilibrium spin state. These relaxation dynamics are expected to vary considerably between different molecules adsorbed to the same surface, due to the various statistical realizations of the phonon dynamics at the different parts of the surface. Molecule-level experiments therefore provide a reason to go beyond the usual ensemble averages of statistical mechanics. In particular, they suggest developing methods for predicting the *probability* of recording particular spin states when random members of a population of high-spin molecules are selected and their spin states are probed with a measuring device[24].

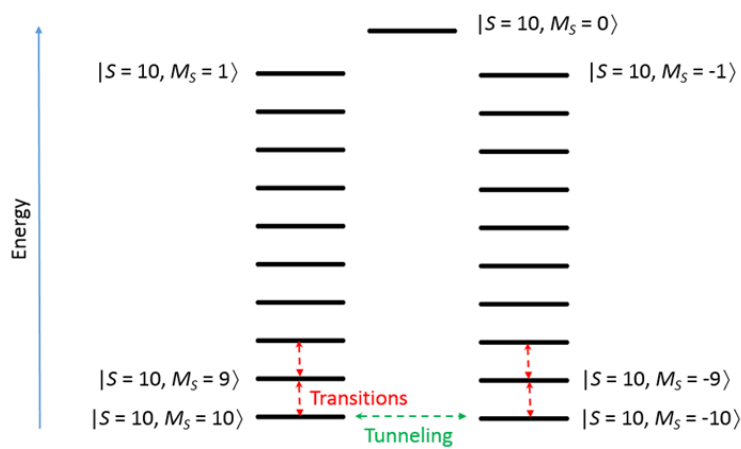

**Figure 1**. Simple sketch of the energy levels of the spin states of a molecule with spin $S = 10$ and internal anisotropy (such as $Mn_{12}$-ac). Only some energy states are labelled to avoid clutter. Transitions that occur within the spin wells are induced by spin-vibron coupling. Transition that occur between spin wells are referred to as spin tunneling.

The case of spin relaxation in high-spin molecules in contact with their environment is a problem from the field of 'open quantum systems'. Open quantum systems are generally inaccessible to first-principles computations, and a great variety of treatments have been developed[25, 26]. Perhaps the oldest such approach to modelling system-environment coupling is through stochastic processes. In this approach, the microscopic details of the environment are ignored, and the coupling is treated as a stochastic modulation to the system[27]. The main advantages of stochastic approaches are that they are relatively simple to implement and simulate. Moreover they give an intuitive description of the system-environment interaction; these are probably the reasons why stochastic approaches remain popular in the literature despite the availability of more detailed methods. For the problem of estimating the probability of measuring particular spin states from a randomly selected molecule adsorbed to a surface, a stochastic theory would aim to calculate the *probability density function* for the coefficients of the spin wave function (or more accurately, of the density matrix elements for the spin state).

To build any stochastic model, we must address the delicate question of which kind of stochastic process is appropriate for the particular problem at hand. For the case of a molecule adsorbed to a surface, in which the complicated molecule-phonon interaction must be considered, the answer to this question is unclear. In this paper, we present the *stochastic Boltzmann equation* (SBE) as a possible formalism for these kinds of problems. The SBE is an equation of motion for the probability density of the density matrix of a system whose dynamics are described by a spin Hamiltonian containing a stochastic term. The stochastic term models the system-environment coupling. While the regular Boltzmann equation from statistical mechanics also describes the time-evolution of a probability density for the state of a system in contact with its environment[28], the SBE is fundamentally different from the regular Boltzmann equation. This difference is due to the stochastic term in the equation of motion of the system in the SBE formalism, whereas in the regular Boltzmann equation formalism the equation of motion for the system is deterministic, and the system-environment interaction is incorporated *ad-hoc* through a 'collision' term that acts on the probability density. As with the regular Boltzmann equation, the time-evolution for the probability density in the SBE is decomposed into two terms. One of these terms arises from the intrinsic dynamics of the system, the other term corrects the first term to account for the system-environment interaction. The latter term is here referred to as a collision term, due to its analogy with the collision term in the regular Boltzmann equation formalism. The value of the SBE approach is that the same collision term can potentially be achieved for a number of different types of stochastic system-environment couplings. This means that the qualitative features of spin relaxation for a number of different physical situations may be studied by solving an SBE with an appropriate collision term under a variety of parameter regimes. The question then is, what kind of collision term is satisfied for a wide variety of stochastic system-environment couplings?

Following treatments of heat transport in solids and rarified gas flow with the regular Boltzmann equation[28, 29, 30], we consider the *relaxation-time approximation* in which the collision term is assumed to be in the form of a first-order loss of probability density. The relaxation-time approximation is expected to be appropriate for any case where the stochastic environment couples weakly to the system. The relaxation-time approximation is rather generic, and a great variety of stochastic processes might be expected to satisfy it, particularly when the system-environment coupling is not too strong. We show that the solutions of the SBE under the relaxation-time approximation can be generated for the case where the system-environment interaction is in the form of discrete, Poisson-distributed collision events (boson gas coupling). This means that we can explore the qualitative features of spin relaxation for any situation which satisfies the relaxation-time approximation by considering boson gas coupling and adjusting the relevant parameters. Moreover, this boson gas coupling is relatively easy to simulate, which means that solutions to the SBE can be built up by simply simulating many realizations of the boson gas coupling, integrating the Schrödinger equation for the system for each realization, and building up a histogram for the density matrix at each time point. The SBE formalism therefore provides information on the spin relaxation dynamics across the whole ensemble of molecules (*via* the probability density for the spin state) and does not resort to ensemble averages, and the ability to solve the SBE *via* simulations of boson gas coupling provides generality in the types of system-environment interaction that can be studied.

In section 2 we will present a stochastic model for spin relaxation in a high-spin molecule and introduce the SBE for this problem. In section 3 we will present a method for solving this SBE under the relaxation-time approximation based on trajectory simulations of the boson gas coupling. Section 4 provides an illustrative example of this approach. Conclusions are given in Section 5.

## 2. Stochastic Boltzmann equation

*i. Stochastic model for magnetic relaxation in high-spin molecules*

Consider a single, high-spin molecule containing $N$ vibrational modes and ground state spin with total spin magnitude $S$. We are interested in the state of the total spin, which we write as

$$|\psi(t)\rangle = \sum_{k=-S}^{S} c_k(t) |S,k\rangle, \qquad (1)$$

where $|S, k\rangle$ refers to a state with $z$-component of spin equal to $M_S = k$. Without loss of generality, we assume that the energies of the states $|S, -S\rangle$, …, $|S, +S\rangle$ are arranged according to the diagram in Figure (1). In the 'giant spin approximation', the Hamiltonian for the state of the total spin can be written as

$$\hat{H}(t) = -D\hat{S}_Z^2 + \hat{T} + \hat{H}_{sv}(t), \qquad (2)$$

where $D$ is the axial anisotropy constant, and $\hat{S}_x$, $\hat{S}_y$, and $\hat{S}_z$ are the $x$-, $y$-, and $z$-components of the spin operator, respectively[1]. The operator $\hat{T}$ is a tunneling operator, defined as

$$\hat{T}|S,k\rangle = \begin{cases} \hbar v_k |S,-k\rangle & k = -S,\ldots,-1,1,\ldots,S \\ 0 & k = 0 \end{cases}, \qquad (3)$$

where the constants $\upsilon_k$ are parameters called tunneling frequencies. The operator (3) allows for total spin to tunnel between adjacent spin states in the two neighboring wells (Figure 1). $\hat{H}_{sv}(t)$ is a linear spin-vibration coupling term,

$$\hat{H}_{sv}(t) = -\sum_{k=1}^{N} Q_k(t) \left[ \sum_{\alpha,\beta=x,y,z} \kappa_k^{\alpha\beta} \hat{S}_\alpha \hat{S}_\beta \right], \qquad (4)$$

where $Q_k(t)$ is the displacement of the $k$th vibrational mode at time $t$ (which is treated as a classical coordinate), and $\kappa_k^{\alpha\beta}$ measures the coupling of the $k$th vibrational mode to the term $\hat{S}_\alpha \hat{S}_\beta$. In real high-spin molecules, the giant spin approximation in equation (2) is expected to be reasonable when exchange interactions between electron spins in the molecule are very strong. The giant spin approximation applies well to the Mn$_{12}$-ac molecule mentioned earlier, in which exchange interactions of the order of 7.3 x 10$^{-3}$ eV between the unpaired electrons of the Mn ions produce a net total spin of $S = 10$ in the spin ground state[1]. For high spin molecules in which coupling between individual electron spins is weaker, one might instead replace the Hamiltonian in (2) with the more general (but comparatively difficult to study) Heisenberg spin model, in which spin states are constructed by directly considering the spin states of the individual electrons in the molecule.

We now couple this molecule to its environment. We suppose that the molecule-environment coupling causes a stochastic modulation to the vibrational mode frequencies. This 'stochastic frequency modulation' approach is used extensively in the field of spectroscopy to model optical relaxation processes[31, 32], and can be described by the Kubo oscillator model[33, 34, 35, 36, 37, 38],

$$Q_k(t) = a_k \exp\left(i \int_0^t W_k(s) ds\right), \tag{5}$$

where $W_k$ is the frequency for the $k$th vibrational mode, which is a stochastic process, and $a_k$ is the amplitude of the mode. The precise nature of the stochastic process $W_k$ can be left unspecified for now. For clarity, the stochastic processes $Q$ and $W$ will be denoted by capital Roman letters, and individual realisations of $Q$ and $W$ will be denoted by small letters $q$ and $w$ respectively (and similarly for other stochastic processes that appear in this paper). Because the Kubo oscillator model does not consider the microscopic details of the environment, it neglects the possibility of a 'back reaction' from the environment to the system, in which the feedback from the system to the environment in turns alters how the environment interacts with the system. Moreover, any quantum aspect of the system-environment interaction is neglected with this approach. We will take up this point further in the conclusions section.

*ii. Stochastic Boltzmann equation*

The density operator for the total spin state of the molecule described in part i is

$$|\phi(t)\rangle\langle\phi(t)| = \sum_{k=-S}^{S}\sum_{j=-S}^{S} C_{jk}(t)|S,k\rangle\langle S,j|, \qquad (6)$$

where $C_{jk}(t) = c_j^*(t) c_k(t)$ and the $c_k(t)$ are the expansion coefficients of the spin wave function given in (1). Under the evolution of the stochastic Hamiltonian in (2), the matrix elements $C_{jk}(t)$ are stochastic processes (for this special case, the $c_k(t)$ are written with small Roman letters, despite being stochastic processes). It is convenient to work in the Liouville space formalism[39], in which the system is represented by the $(2S+1)^2$-dimensional vector

$$\mathbf{R}(t) = \left(C_{-S,-S}(t),\ldots,C_{-S,S}(t), C_{-S+1,-S}(t),\ldots,C_{-S+1,S}(t),\ldots,C_{S,-S}(t),\ldots,C_{S,S}(t)\right).$$

(7)

$\mathbf{R}(t)$ is called the Liouville space vector for the system. Note that Liouville space vectors are contained in a Hilbert space[39]. Let $f(\mathbf{\rho},t)$ denote the probability density function of $\mathbf{R}(t)$. If we randomly choose one molecule from a non-interacting ensemble of molecules, then $f(\mathbf{\rho},t)\delta\mathbf{\rho}$ is the probability that $\mathbf{R}(t)$ is contained in a region of volume $\delta\mathbf{\rho}$ centered at point $\mathbf{\rho}$ in the Hilbert space. This *statistical probability* must be distinguished from the *quantum probability* $C_{kk}(t)$ that the system will collapse into spin state $|S,k\rangle$ when the spin state at time $t$ is probed by a measuring device.

The stochastic Boltzmann equation (SBE) is an equation of motion for the probability density $f(\boldsymbol{\rho},t)$. In the Appendix, the SBE for the system described in section *i* is shown to be

$$\frac{\partial f(\boldsymbol{\rho},t)}{\partial t} + \frac{1}{i\hbar}\left\langle \hat{L}^{\pi}(t)\boldsymbol{\rho} \cdot \frac{\partial f(\boldsymbol{\rho},t|\pi)}{\partial \boldsymbol{\rho}} \right\rangle = \left.\frac{\partial f(\boldsymbol{\rho},t)}{\partial t}\right|_{\text{coll}}. \tag{8}$$

The second and third terms are interpreted as follows.

*Second term.* For simplicity, let us consider the case of a molecule with $N = 1$ vibrational mode. With reference to equation (5) for the vibrational mode amplitude, let us suppose that the path for the vibrational frequency from time 0 to time $t$ and amplitude at time 0 is not stochastic but instead fixed to one path realization $\pi$. The notation $f(\boldsymbol{\rho},t|\pi)$ denotes the probability density for $\mathbf{R}(t)$ computed under this fixed path (i.e., $f(\boldsymbol{\rho},t|\pi)$ is a conditional probability density). The probability described by $f(\boldsymbol{\rho},t|\pi)$ arises because the initial condition $\mathbf{R}(0)$ is not specified precisely and is a random variable. The super-operator $\hat{L}^{\pi}(t)$ has elements

$$L^{\pi}_{jk,mn}(t) = H^{\pi}_{jm}(t)\delta_{kn} - H^{\pi *}_{kn}(t)\delta_{jm}, \tag{9}$$

where the Hamiltonian matrix elements refer to the Hamiltonian in equation (2), but with the stochastic spin-vibron coupling term replaced with

$$\hat{H}_{sv}^{\pi}(t) = -\sum_{k=1}^{N}\left[\sum_{\alpha,\beta=x,y,z}\kappa_k^{\alpha\beta}\hat{S}_\alpha\hat{S}_\beta\right]q_k(0)\exp\left(i\int_0^t w_k(r)dr\right). \quad (10)$$

Equation (10) contains no stochastic component (unlike equation (5)), because the path for the frequency $w_k$ of the vibrational mode between times 0 and $t$, and also the vibrational mode amplitudes $q_k$ at time 0, are fixed to the values given by $\pi$. The notation $\langle\ \rangle$ in equation (8) denotes a statistical average over all possible realisations $\pi$ of the paths for the vibrational mode and amplitudes up to time $t$. The second term in equation (8) can be roughly interpreted as the rate of flow of probability into a small region centered at point $\boldsymbol{\rho}$ due to the *intrinsic quantum dynamics* of the system, averaged over all possible realisations of the paths for the frequency process. The case of a molecule with $N > 1$ vibrational modes is interpreted similarly, but this time $\pi$ is a set of $N$ realisations for the paths of the $N$ frequency processes.

*Third term.* The 'collision' term $\partial f(\boldsymbol{\rho},t)/\partial t\big|_{coll}$ is a correction for the second term to account for the change in probability density at point $\boldsymbol{\rho}$ at time $t$ due to the stochastic interaction with the environment. It does not appear possible to provide a systematic method for calculating the collision term for an arbitrary stochastic frequency process. We will instead make the *relaxation-time approximation*,

$$\left. \frac{\partial f(\boldsymbol{\rho},t)}{\partial t} \right|_{coll} = \left\langle \upsilon_\pi f(\boldsymbol{\rho},t|\pi) \right\rangle, \tag{11}$$

where $\upsilon_\pi^{-1}$ is a constant called the *relaxation time* for the path $\pi$. Equation (11) be can understood by substituting it into the SBE in equation (8), which leads to

$$\frac{\partial f(\boldsymbol{\rho},t)}{\partial t} + \left\langle \frac{1}{i\hbar} \hat{L}^\pi(t)\boldsymbol{\rho} \cdot \frac{\partial f(\boldsymbol{\rho},t|\pi)}{\partial \boldsymbol{\rho}} - \upsilon_\pi f(\boldsymbol{\rho},t|\pi) \right\rangle = 0. \tag{12}$$

For a fixed set of frequency paths $\pi$, the first-term inside the square brackets describes the flow of probability density into the region around point $\boldsymbol{\rho}$ at time $t$ due to the intrinsic quantum dynamics of the system. Then, the second term $-\upsilon_\pi f(\boldsymbol{\rho},t|\pi)$ says that there is a 'leakage' of this probability density from this region with time-constant $\upsilon_\pi^{-1}$ due to the stochastic interaction with the environment at time *t*. In practice, we will assume that the constants $\upsilon_\pi^{-1}$ only depend upon the state of the path $\pi$ at time *t*, rather than on the entire history of the path.

The formal mathematical difference between equation (12) and the regular Boltzmann equation is the averaging over the paths $\pi$ of the stochastic frequency process. This difference arises as follows. In the SBE approach, the effect of the environment appears explicitly through the stochastic terms in the Hamiltonian in (5). On the other hand, in the regular Boltzmann equation approach, the effect of the environment does not explicitly appear in the equations of motion for the system themselves, and is included *ad-hoc* by the addition of a collision term to the equation of motion for the probability density function[28]. Because the effect of the environment appears explicitly in the equations of motion for the current problem, we need to average over the paths of the normal mode frequencies in order to follow the steps of the derivation of the regular Boltzmann equation (see the Appendix). Both the SBE and the regular Boltzmann equation describe the same phenomenon (time evolution of the probability density of the state of the system), however the SBE is mathematically more sensible for the case where stochastic processes explicitly appear in the Hamiltonian for the system. The formal differences between the SBE and the regular Boltzmann equation mean that the numerous techniques that have been developed to solve the latter are not expected to apply to the SBE, because in this case we must average over all distinct paths of the stochastic frequency processes.

In passing, note that another type of 'stochastic Boltzmann equation' is discussed in the fluctuating hydrodynamics literature[40, 41], however in this case 'stochastic' refers to the presence of stochastic processes in the collision term of the regular Boltzmann equation, rather than stochastic processes in the equations of motion for the systems in the ensemble.

### 3. Boson Gas Coupling

Even if we are satisfied working with the relaxation-time approximation, integrating the SBE directly appears to be very difficult due to the need to average over all realisations of the paths of the frequency processes. To proceed, we consider a Monte Carlo approach that is used to solve the regular Boltzmann equation for the case of rarefied gas dynamics, in which a sample of gas particles undergoing a stochastic motion are simulated, and the the laws describing this stochastic motion are chosen such that the solution to the Boltzmann equation is satisfied[42, 43]. A similar approach has also been used in the field of heat transport in solids[44]. In the present case of the SBE, we must identify a generic stochastic frequency process that satisfies the relaxation-time approximation in equation (11) and is easy to simulate.

We can identify such a process by supposing that the environment surrounding the molecule is in the form of a 'boson gas'. Let us suppose that for each vibrational mode in the molecule the frequency is of the form

$$W_k(t) = \hbar \omega_k \left( V_k(t) + 1/2 \right), \tag{13}$$

where $V_k(t)$ is a non-negative, integer-valued stochastic process. The boson gas has an energy density $d(u)$, where $u$ is a boson frequency. Now, suppose that bosons from the density $d(u)$ collide independently with the molecule at Poisson-distributed random times. In particular, for the normal mode with frequency $\omega_k$, bosons of frequency $u = \omega_k$ collide with the molecule at an average rate $\lambda_k$, where

$$\lambda_k = B(\omega_k) d(\omega_k), \tag{14}$$

where $B(\omega_k)$ is an $\omega_k$-dependent proportionality constant. When a boson of frequency $\omega_k$ collides with the molecule, the stochastic process $V_k$ either increases by one unit (stimulated absorption) or decreases by one unit (stimulated emission) with equal probability. Note that equation (14) satisfies Einstein's radiation laws for the case of a photon gas. For each vibrational mode *k*, we also assume that $V_k$ decreases by one unit (*spontaneous emission*) with an average rate of $\gamma_k$ at Poisson-distributed times. $\gamma_k$ is given by Einstein's formula for spontaneous emission,

$$\gamma_k = \lambda_k \left( e^{2\pi\hbar\omega_k/k_B T} - 1 \right). \tag{15}$$

Spontaneous emissions are assumed to occur independently of stimulated emissions and absorptions. Moreover, spontaneous and stimulated emission processes are not allowed to occur if $V_k = 0$. The inclusion of spontaneous emission processes is necessary to ensure that the vibrational modes can reach thermal equilibrium with the environment, and that conservation of energy (system + environment) is achieved. Under these assumptions, we can show that equation (11) is satisfied with (see Appendix 2)

$$\upsilon_\pi = \sum_{k=1}^{N} (2\lambda_k + \gamma_k). \tag{16}$$

In equation (16), $\upsilon_\pi$ is interpreted as follows. Consider a molecule from the ensemble whose stochastic frequencies follow path $\pi$ up to time *t*. Then the relaxation-time $1/\upsilon_\pi$ is the average length of time that we have to wait until the next molecule-environment interaction (stimulated emission, absorption, or spontaneous emission) occurs. Note that for any path $\pi$ in which the frequency $W_k$ has value 0 at time $t$, the term $(2\lambda_k + \gamma_k)$ in equation (15) is set to $\lambda_k$ to exclude spontaneous and stimulated emission processes. The parameters $\upsilon_\pi$ are therefore time-dependent, as mentioned at the end of the previous section, but do not depend upon the entire history of the paths up to time $t$.

Stochastic frequency processes stimulated according to the above scheme lead to probability densities that satisfy the SBE in equations (8) under the relaxation-time approximation. We refer to this scheme as the *boson gas coupling* approach. To estimate the probability density for any problem which satisfies the relaxation-time approximation (regardless of whether the actual environment is a boson gas or not), we generate $n$ independent samples of $N$ trajectories for the stochastic frequency processes by following the rules above, and integrate the Schrodinger equation for each of the $n$ samples up to a desired time. The Liouville space vector $\mathbf{R}(t)$ is then built by taking appropriate products of the real and imaginary parts of the coefficients estimated from the Schrodinger equation for each sample. Assuming large enough $n$, we can then visualize the probability distribution of the elements of $\mathbf{R}(t)$ by plotting a histogram of the Liouville space vectors estimated from each sample. By repeating this processes for a variety of values of the parameter $B(\omega_k)$ and the parameters for $\rho(\omega_k)$, we can explore the qualitative features of any solution to the SBE under the relaxation time approximation. We will return to this point in section 5.

## 4. Example calculation

To illustrate the application of the SBE approach, we perform a simple calculation for the case of a molecule with spin states arranged according to Figure 1 undergoing spin relaxation due to stochastic coupling with an environment and also tunneling between spin states of adjacent wells. We will explicitly assume that the environment is a gas of infrared photons, i.e., a boson gas as described in the previous section with an energy density $d(u)$ weighted heavily in the infrared region. Because the probability density in (8) is a complicated function of $(2S+1)^2 = 441$ elements of the Liouville space vector $\boldsymbol{\rho}$ plus one time variable, we will instead compute the two-dimensional marginal distributions

$$f\left(c_{i,j}, t\right) = \int f(\boldsymbol{\rho}, t) d^{ij} c \qquad (17)$$

for several interesting elements $c_{i,j} = c_i^* c_j$ of $\boldsymbol{\rho}$, where the notation $d^{ij}c$ means that the integral in (17) runs over all variables of $\boldsymbol{\rho}$ except for $c_{i,j}$. We will consider a case where the photon frequency distribution $d(u)$ is a Gaussian function of $u$, namely

$$d(u) = A \exp\left(-(u-u_0)/2b^2\right), \qquad (18)$$

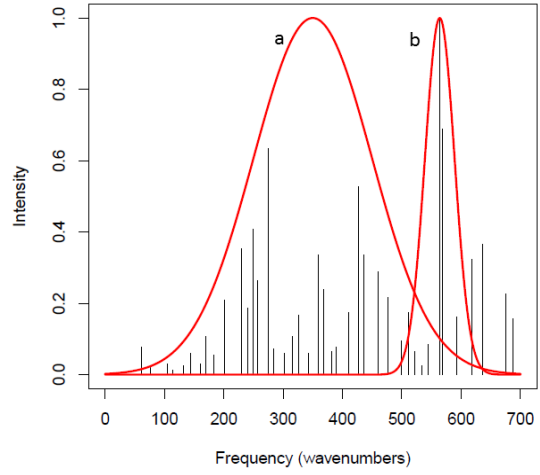

Figure 2. Vertical lines show the infrared spectrum of $Mn_{12}$-ac (from reference 45). The curves show the photon frequency distribution $d(u)$ for the broad Gaussian case (a) and narrow Gaussian case (b).

where $u_0$ is the center of the frequency distribution, $b$ is its bandwidth, and $A$ is an intensity parameter. This situation is achievable under laser irradiation. For the parameter $B(\omega_k)$, we will choose the usual value used in Einstein's radiation theory, namely

$$B(\omega_k) = \frac{e_0^2 \mu_k}{6\varepsilon_0 \hbar^2} \qquad (19)$$

where $e_0$ is the electron charge, $\varepsilon_0$ is the permittivity of vacuum, and $\mu_k$ is the transition dipole moment. We further set $\mu_k = l^2 I(\omega_k)$, where $I(\omega_k)$ is the intensity of the absorption at frequency $\omega_k$ for the vibrational mode $k$, and $l$ is a 'distance' parameter which measures how far charge shifts in the molecule when a photon of frequency $\omega_k$ is absorbed by vibrational mode *k*. The vibrational frequencies $\omega_k$ and intensities $I(\omega_k)$ were taken from an IR spectrum for the single-molecule magnet Mn$_{12}$-ac calculated by Pederson *et al* [45]. Only the vibrational frequencies for the 41 clear peaks in this data were considered (see Figure 2). The spin-vibrational coupling parameters $\kappa_k^{\alpha\beta}$ were each taken to be

$$\kappa_k^{\alpha\beta} = \theta_k I_0(\omega_k), \qquad (20)$$

where $\theta_k$ is a coupling constant which is independent of $\alpha, \beta$ and $I_0(\omega_k)$ is the contribution of the manganese atoms to the infrared intensity at frequency $\omega_k$. These parameters $I_0(\omega_k)$ were also taken from the calculations of Pederson *et al*[45]. All parameters used in this calculation are summarized in Table 1. These coupling parameters for Mn12-ac are chosen out of convenience, and the following calculations are *not* meant to represent any particular real molecule. According to Table 1, we are considering a case of very strong tunneling between wells and strong coupling to the photon gas environment, which, while unrealistic, is necessary to observe interesting relaxation dynamics over a short time scale. We will distinguish between the *stochastic* contribution to spin relaxation (relaxation due to the collision term on the right-hand side of the SBE in equation (8)) and the *tunneling* contribution to spin relaxation (due to quantum dynamical transitions between states described by the second term on the left-hand side of equation (8)).

|  |  | Reference |
|---|---|---|
| Time step | 0.5 fs |  |
| Number of simulations | 1000 |  |
| Integration method | 4th-order Runge Kutta |  |
| $N$ (number of vibrational modes) | 41 | 45 |
| $\omega_k$ | (See Fig. 2) | 45 |
| $D$ | -0.46 cm$^{-1}$ | 1 |
| $a_k$ | 0.1 Å for all $k$. |  |
| $\upsilon_k$ | 0.25 ps$^{-1}$ for all $k$ |  |
| Temperature | 300 K |  |
| $A$ | 0.5 |  |
| $l$ | 0.1 Å |  |
| $\theta_k$ | 2.0 for all $k$ |  |
| $v_0$, $b$ | 350 cm$^{-1}$, 100 cm$^{-1}$ ('broad Gaussian case') 564 cm$^{-1}$, 25 cm$^{-1}$ ('narrow Gaussian case') |  |

Table 1. Parameters used in the calculations of section 4.

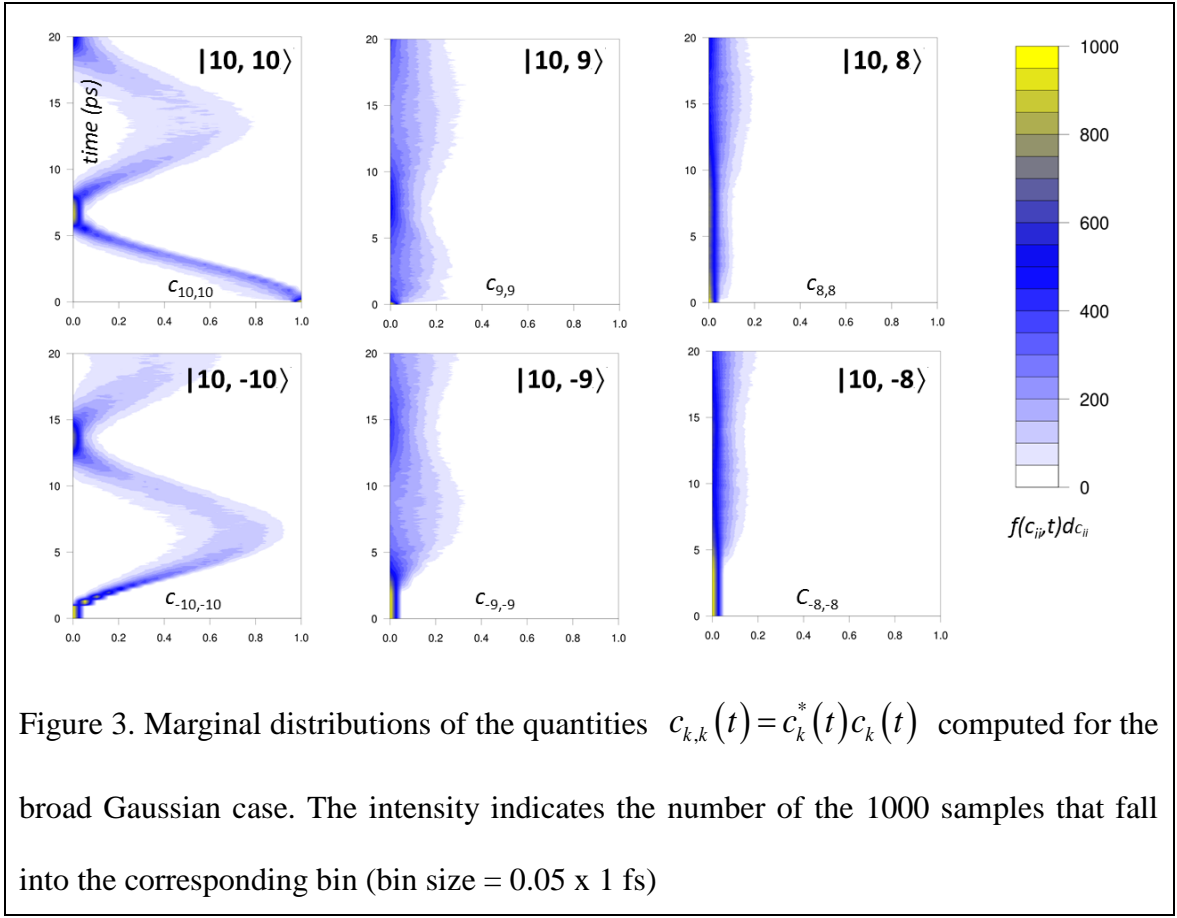

Figure 3. Marginal distributions of the quantities $c_{k,k}(t) = c_k^*(t) c_k(t)$ computed for the broad Gaussian case. The intensity indicates the number of the 1000 samples that fall into the corresponding bin (bin size = 0.05 x 1 fs)

We first consider the case of an molecule in a gas of IR photons with a Gaussian frequency distribution centered at 350 cm$^{-1}$ with a bandwidth of 100 cm$^{-1}$ (broad Gaussian case). As shown in Figure 2, this distribution covers a majority of the vibrational modes in the molecule. We first discuss the distribution $f(c_{k,k}, t)$ of the elements of the Liouville space vector $\mathbf{R}(t)$ of the form $C_{kk}(t)$. Thus, $f(c_{k,k}, t) \delta c_{k,k}$ is the statistical probability that, if we choose a molecule randomly from an ensemble of molecules, the quantity $C_{kk}(t)$ will be located within a small region of width $\delta c_{k,k}$ about $c_{k,k}$. Then, this $c_{k,k}$ corresponds to the quantum mechanical probability this molecule will collapse into the spin state $|S,k\rangle$ during an attempt to measure its spin state. The probability that a randomly chosen molecule from an ensemble is measured in spin state $k$ is therefore $f(c_{k,k},t)\delta c_{k,k} \times c_{k,k}$. Figure 3 plots the probability densities for the cases $k = 10, 9, 8, -10, -9, -8$, as estimated by the approach described above. The most obvious feature of the results in Figure 3 is the oscillatory shifting of the probability density between states $|S,10\rangle$ and $|S,-10\rangle$ due to strong spin tunneling between states in neighboring wells. This oscillatory shifting is generated by the second term on the left-hand side of equation (8), which arises from the intrinsic quantum dynamics of the system. It can be seen that the probability densities become relatively diffuse at the turning points of the oscillations, which is due to a 'leaching' of the probability density into states 9 and -9 due to the stochastic interactions with the surrounding boson gas environment. The occurrence of probability density leaching at the oscillation turning points does not appear to have been mentioned in other theoretical studies of spin relaxation in high-spin molecules, and cannot be observed without simulating the probability density for the density operator. This leaching occurs under the influence of the 'collision' terms on the right-hand side of equation (8), which is linear in the amount of density localised in a particular state. These calculations suggest that the stochastic contribution to spin relaxation is only significant when the spin is localised to one well.

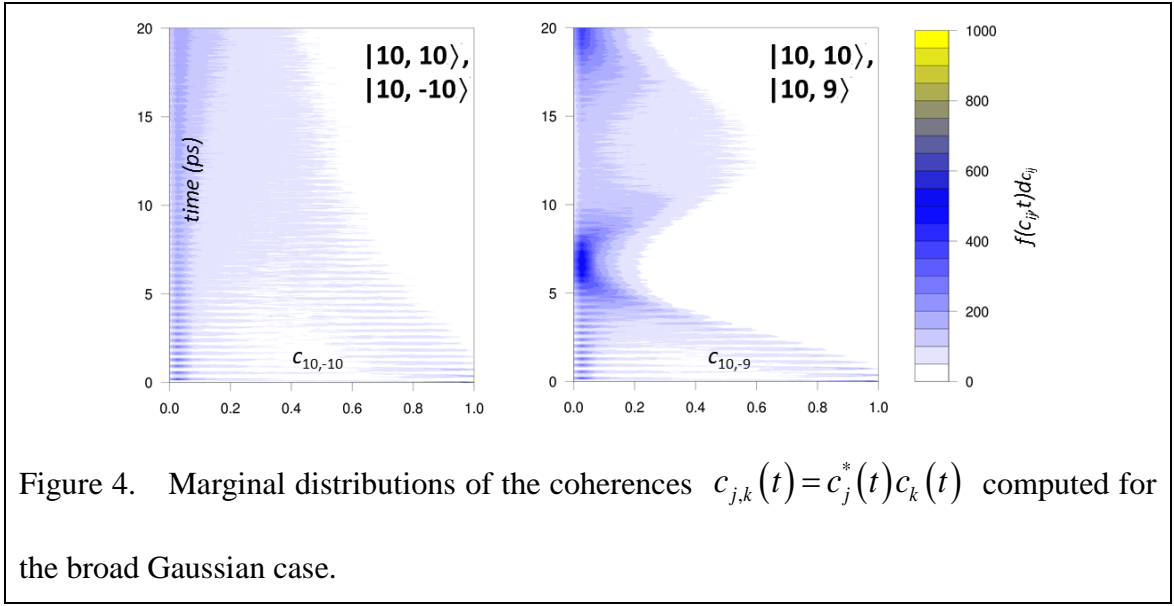

Figure 4. Marginal distributions of the coherences $c_{j,k}(t) = c_j^*(t) c_k(t)$ computed for the broad Gaussian case.

We now consider the probability density $f(c_{n,m}, t)$ of the elements of the Liouville space vector $\mathbf{R}(t)$ of the form $C_{nm}(t)$, where $n \neq m$. Consider two states $|S, n\rangle$ and $|S, m\rangle$ which contribute to the total spin state of a molecule in the ensemble. The complex number $C_{nm}(t)$ is called the 'coherence' between states $|S, n\rangle$ and $|S, m\rangle$, and is an oscillating quantity whose frequency depends upon the relative phase of the two states $|S, n\rangle$ and $|S, m\rangle$ at time *t*. Then, $f(c_{n,m}, t) \delta c_{n,m}$ gives the probability that the molecule randomly selected from an ensemble will have a total spin in which states $|S, n\rangle$ and $|S, m\rangle$ have a coherence contained in a region of size $\delta c_{n,m}$ centered at point $c_{n,m}$ in the complex plane. Figure 4 plots the probability density of the real part of the coherence for states $|S, 10\rangle$ and $|S, -10\rangle$ and for states $|S, 10\rangle$ and $|S, 9\rangle$ (= $C_{10,-10}(t)$ and $C_{10,9}(t)$, respectively). The density of the coherence $C_{10,-10}(t)$ shows a clear oscillatory structure up until about 5 ps, after which it quickly becomes lost. This coincides with the onset of the first turning point in the density $f(c_{-10,-10},t)$ for the state $|S,-10\rangle$ shown in Figure 2 and the subsequent leaching of this density into the state $|S,-9\rangle$ due to the stochastic component of spin relaxation. A similar result is seen for the coherence $C_{10,9}(t)$, in which the oscillatory structure is lost after about 10 ps, coinciding with the occurrence of the turning point for the density $f(c_{10,10},t)$ for the state $|S,10\rangle$ (all coherences with $|M_S|\leq 8$ showed essentially similar behavior to $C_{10,9}(t)$). These results suggests that phase coherence between spin states is rapidly lost when the spin is localised to a single well and the stochastic component to spin relaxation becomes important. This information is useful because the loss of phase coherence would inhibit the ability to observe spin oscillations due to tunneling when measurements are averaged over an ensemble of molecules. Thus, by engineering molecules to have large tunneling frequencies, we would expect for the stochastic component for spin relaxation to be reduced, which in turn may decrease the rate of phase decoherence between different states.

We now consider the 'narrow Gaussian case' where the frequency distribution of the IR photons is centered at 564 cm$^{-1}$ with bandwidth of 25 cm$^{-1}$. As shown by Figure 2, this distribution is narrowly focused on the vibrational mode with the strongest absorption in the IR field. Figure 5 plots the densities $f(c_{k,k}, t)$ for the cases $k$ = 10, -10, 9, -9, 8, and – 8. The evolution of the probability densities is qualitatively similar to the case of a broad Gaussian frequency distribution centered at 350 cm$^{-1}$ shown in Figure 3. However, the turning points for the $|S, 10\rangle$ and $|S, -10\rangle$ states for the present case occur slightly later than for the case of the broad Gaussian frequency distribution. For example, for the present case most of the probability density for the state $|S, -10\rangle$ returns to $c_{-10,-10} = 0$ for the first time at about 15 ps, which is about 2 ps later than for the case of a broad distribution. This delay in the onset of the turning point can only be due to more probability density being leached into the lower spins states due to shorter relaxation times in the collision term for the present case compared to the case of a broad Gaussian distribution of frequencies. Indeed, the probability density at the turning points for the present case appears to be slightly more diffuse than for the broad Gaussian case, confirming that more leaching of probability density into the lower spin states is occurring for this case. The behavior for the distribution of the coherences $C_{n,m}(t)$ for the present case was essentially the same as what was seen for the case of a broad Gaussian distribution of frequencies (results not shown). One conclusion is that the oscillation of the probability density between adjacent spin states due the quantum dynamical term in equation (8) is modulated by the leaching of the probability density due to the collision term. A second conclusion is that, if the frequency distribution of the boson gas is tuned to vibrational modes with strong coupling to the IR field and the spin state, then spin relaxation of the molecule can be more dramatic than when the frequency distribution is broader and spread over many vibrational modes. This suggests that the stochastic contribution to the spin relaxation dynamics could be tuned by adjusting the vibrational mode frequencies of the molecule.

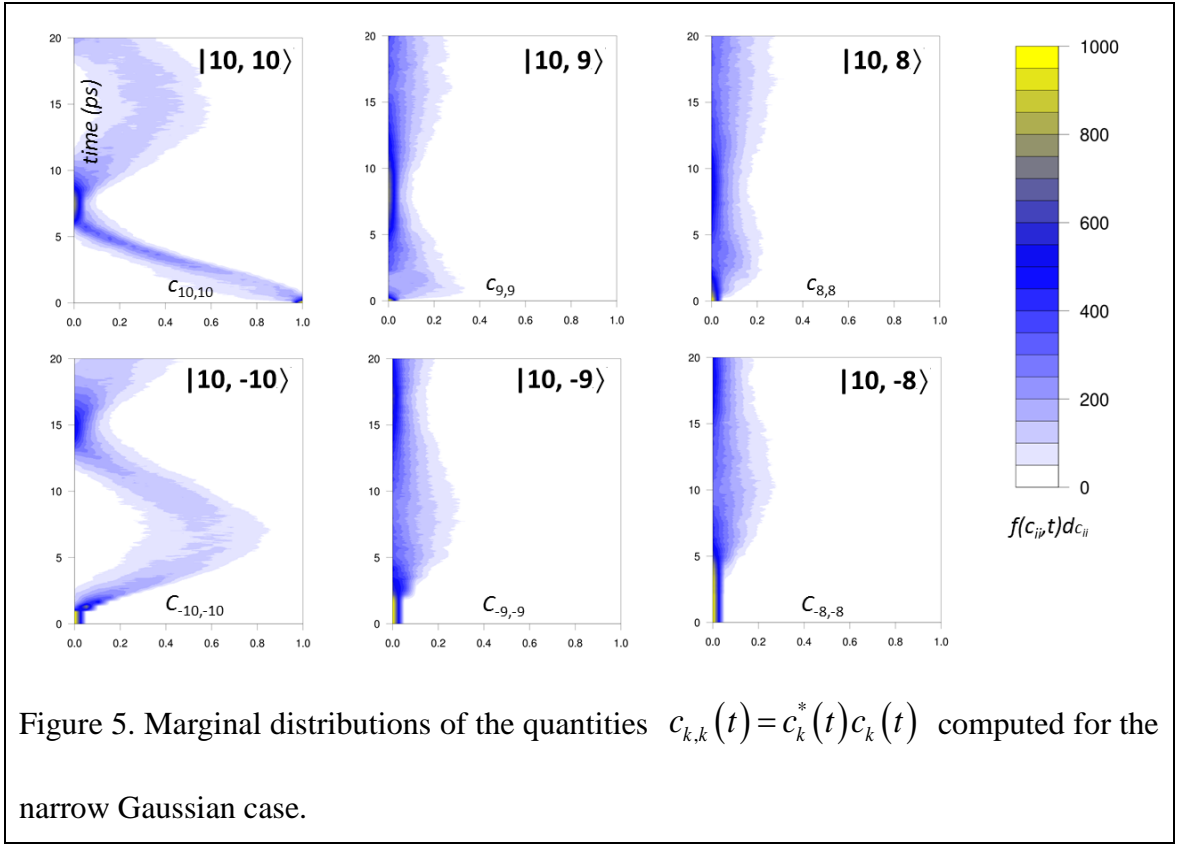

Figure 5. Marginal distributions of the quantities $c_{k,k}(t) = c_k^*(t) c_k(t)$ computed for the narrow Gaussian case.

While the results here were calculated by assuming a photon gas environment, they are expected to be generally relevant for all cases in which the stochastic coupling between the molecule and surroundings satisfies the relaxation-time approximation, as highlighted above.

## 5. Conclusions

This paper has introduced the stochastic Boltzmann equation (SBE) as a method for exploring the spin dynamics of high-spin molecules coupled to a stochastic environment. The SBE is a time-evolution equation for the probability density of the spin density matrix of the system, in which the spin Hamiltonian contains a stochastic term representing coupling with the external environment. The value of this formalism is that probability densities are the relevant quantity to describe experiments performed on individual, non-interacting, molecules (such as single-molecule magnets adsorbed to a surface in a scanning tunneling microscopy experiment[3, 22]. By analogy with treatments of heat transport with the regular Boltzmann equation[28], we proposed a *relaxation time approximation*, in which the loss of probability density due to the stochastic interaction with the environment is linear in the probability density. We showed that under the relaxation-time approximation the SBE can be solved by performing simple trajectory simulations for the case of a system coupled to a 'boson gas' environment. This provides a useful means to explore the solution space of the SBE under the relaxation time approximation by simply solving the SBE for a variety of different parameter regimes for the boson gas coupling. It would therefore be of great interest to investigate whether the case of a high-spin molecule adsorbed to a metal surface can be treated within the relaxation-time approximation, because then the SBE approach might be used to study this important (but otherwise formidable) problem. As with all stochastic models, the SBE assumes a classical system-environment interaction. Sophisticated computational techniques such as the hierarchy equations of motion are required to properly quantify the contribution of quantum mechanical effects on the system-environment interaction for realistic systems[46], and it is beyond the scope of this paper to explore them here. In a study of quantum entanglement between a single harmonic oscillator weakly coupled to a bath of harmonic oscillators, the system-bath entanglement was found to vanish entirely at about 5 K, and quantum mechanical contributions to the system-bath interaction were very small at temperatures above about 10 K[47]. We might therefore expect that the present SBE approach, which assumes classical coupling between the vibrational modes of the molecule and phonons of the environment, is reasonable for studying the dynamics of real molecules at temperatures on the order of 10 K and above.

While further work is necessary to address the validity of the relaxation time approximation used here, we can gain some insight into its physical meaning by considering the case of a boson gas environment. At the end of Appendix 2, it is shown that when the occurrence of vibrational transitions within the molecule becomes very frequent on the time scale over which the system evolves, then the relaxation time approximation breaks down. This suggests that the relaxation time approximation applies when the occurrence vibrational transitions within the molecule due to the action of the surroundings is very infrequent. Experimentally, this situation could be achieved by either altering the environment (e.g., lowering the temperature) or altering the molecule in order to weaken the relevant vibrational-environment couplings. For the latter possibility, let us consider the case of a molecule such as $Mn_{12}$-ac, in which the spin arises from interactions between the transition metal ions within the molecule. In the case of $Mn_{12}$-ac, these metal ions are 'buried' by a bulky organic framework made up of acetate ligands, and are often viewed as being 'protected' by the environment by these ligands. Thus, if the molecule is structured so that the vibrational modes that couple strongly to the environment are only weakly coupled to the molecule's spin centers, then the relaxation time approximation might be expected to hold. For the case of high-spin molecules derived from $Mn_{12}$-ac, the validity of the relaxation time approximation might therefore be adjusted by altering chemical the structure of the ligands surrounding the metal centers. Such a study would be a useful exercise in synthetic organic chemistry, and could be supplemented with quantum chemical calculations to determine how molecular structure affects which vibrational modes couple most strongly to the metallic spin centers.

In order to illustrate the SBE approach, we presented some calculations for the spin relaxation in an artificial molecule and explored some general features of spin relaxation due to spin tunneling between wells and stochastic interactions with the environment. The second-term on the left-hand side of the SBE in equation (8) gives the contribution to spin-relaxation from the intrinsic quantum dynamics of the system (spin tunneling between wells), and the collision term on the right-hand side gives the contribution from the stochastic interaction between the system and the surrounding environment. The key result obtained from these calculations is that the contribution of the collision term to spin relaxation is particularly significant when the spin is localised to a single well, however is relatively unimportant when the spin is delocalised between wells. This new insight suggests that the spin relaxation process can be visualized in terms of a probability density oscillating between adjacent spin states of the two wells, with the density 'leaking' into neighboring states within the same well at the turning points of the oscillation. Note that this insight can only be acquired by approaches such as the SBE, in which the probability density for the total spin state is calculated directly. Calculations using more realistic parameters are necessary to 'map out' the entire spin relaxation process for real molecules over longer periods of time. Using realistic parameters for e.g., a $Mn_{12}$-ac molecule, the entire relaxation process is expected to involve at least microsecond time-scales even at relatively high ( > 100 K) temperatures[20]. However, calculations over long time periods using realistic parameters lead to a new problem: on the microsecond time-scale, these sub-picosecond oscillations of the vibrational modes are extremely fast, requiring special integration techniques for the numerical simulation (such as adaptive-step algorithms which can properly account for the stochastic trajectories). This will be the target of future work.

There are some points to note about the stochastic 'boson gas' simulation method for solving the SBE. The idea of solving the regular Boltzmann equation for rarified gas dynamics and heat transport in solids by stochastic trajectory simulation of gas particles is an active field with a long history[43]. In this research, it has been pointed out that the solutions obtained by such simulations do not converge in a mathematically strong sense ('convergence in probability') to solutions to the regular Boltzmann equation as the number of simulated particles tends to infinity, but rather in the mathematically weaker sense of 'convergence in distribution', and so special techniques have been developed to deal with this problem[42]. Based on comparisons between these Monte Carlo approaches to solving the regular Boltzmann equation for rarified gas dynamics, and our boson gas simulation method for solving the SBE for molecules adsorbed to a surface, we might expect that the solutions presented in section 4 are only such 'weak' solutions to the SBE. In practical terms, this means that the moments of the distributions in section 4 should be good approximations to the moments of the solutions to the SBE, however the detailed shape of the results may possible be different. This presents an interesting opportunity for further research, however it is beyond the scope of the present paper to investigate this mathematical problem. It is also noteworthy that it is much easier to construct such a Monte Carlo scheme for the case of the SBE compared to the case of rarified gas dynamics in the regular Boltzmann equation, because in the present case the molecules of the ensemble are not interacting with each other, but rather with an external environment whose details are not of interest.

Another interesting possibility for future research with the SBE is the following. Throughout this paper, we have emphasized that the 'general, qualitative features' of spin relaxation under the relaxation-time approximation can be studied by solving the SBE for the case of a boson gas coupling with various choices of parameters for the boson gas. But is it possible to quantitatively study the spin relaxation for a molecule adsorbed to a metal surface, by fitting the boson gas coupling parameters to the actual surface-molecule coupling (assuming that the latter falls within the relaxation-time approximation)? This approach resembles how some exchange-correlation functionals are formulated for density functional theory[48]. For the present case, we might naively consider performing many atomistic calculations to generate the probability density of the Liouville space vector in (7), and attempt to fit equation (11) at each time point to a set of relaxation times $\upsilon_\pi$. If the fit is sufficient, then one might be able to extract the parameters of the Boson gas algorithm from $\upsilon_\pi$. We will consider this further as the next step for this research, in addition to the other problems mentioned above.

**Appendix 1. Derivation of the Stochastic Boltzmann Equation**

We will derive the stochastic Boltzmann equation with reference to the model presented in section 2 part *i*. The time-evolution of $\mathbf{R}(t)$ (equation (7) of the main text) is given by the Liouville equation

$$\frac{\partial \mathbf{R}(t)}{\partial t} = -\frac{i}{\hbar}\hat{L}(t)\mathbf{R}(t), \tag{A1}$$

where $\hat{L}$ is the tetradic operator with elements[39]

$$L_{jk,mn}(t) = H_{jm}(t)\delta_{kn} - H_{kn}^{*}(t)\delta_{jm}, \tag{A2}$$

and we are referring to the Hamiltonian in equation (2). Under the evolution described by equation (A1), $\mathbf{R}(t)$ is a stochastic process. For the purposes of this derivation, let us consider an ensemble consisting of a large number of non-interacting copies of this molecule, each of which is subject to an independent realisation of the stochastic frequency modulation. Then $f(\boldsymbol{\rho},t)\delta\boldsymbol{\rho}$ is the fraction of the systems with Liouville state vector contained within a region of volume $\delta\boldsymbol{\rho}$ centered at point $\boldsymbol{\rho}$ in the Hilbert space where the Liouville state vectors belong. Moreover, $f(\boldsymbol{\rho},t)$ is the probability density function for the Liouville state vector $\mathbf{R}(t)$ for the system. To avoid clutter, we assume that the number of vibrational modes *N* is equal to 1. The extension to the general case follows the same steps as reported here.

Applying the rule of total probability to the density function $f(\boldsymbol{\rho},t)$, we can write

$$f(\boldsymbol{\rho},t) = \langle f(\boldsymbol{\rho},t|\pi) \rangle, \qquad (A3)$$

where $\pi$ denotes the path of the stochastic process $W$ up to time $t$ and the angular brackets indicates expectation with respect to a probability measure on the path space. The conditional probability density $f(\boldsymbol{\rho},t|\pi)$ is interpreted as follows. Consider only the systems in the ensemble whose frequency paths up to time $t$ are $\pi$, and call this a subset of systems $S$. Then $f(\boldsymbol{\rho},t|\pi)\delta\boldsymbol{\rho}$ is the fraction of systems in $S$ that have their Liouville space vectors contained in a region of size $\delta\boldsymbol{\rho}$ centered at point $\boldsymbol{\rho}$ at time $t$. Starting from equation (A3), we can derive the SBE by following the usual steps for the regular Boltzmann equation given the frequency path $\pi$ [28]. Consider a short time interval from $t$ to $t+\Delta t$. If no stochastic modulation occurs in this interval to any of the molecules in $S$, then the number of molecules $f(\boldsymbol{\rho},t|\pi)\delta\boldsymbol{\rho}$ follows the conservation rule

$$f(\boldsymbol{\rho},t|\pi)\delta\boldsymbol{\rho} = f(\boldsymbol{\rho}+\Delta\boldsymbol{\rho},t+\Delta t|\pi)\delta\boldsymbol{\rho}, \qquad (A4)$$

where $\Delta\boldsymbol{\rho}$ is the change in Liouville state vector $\boldsymbol{\rho}$ from time $t$ to time $t+\Delta t$ for a molecule in $S_k$. According to equation (A1), this $\Delta\boldsymbol{\rho}$ is given by

$$\Delta \boldsymbol{\rho} = -\frac{i}{\hbar} \hat{L}^\pi \boldsymbol{\rho} \Delta t, \tag{A5}$$

where $\hat{L}^\pi(t)$ is defined as in (A2), but with the stochastic spin-vibrational coupling term in the Hamiltonian replaced with the non-stochastic coupling term

$$\hat{H}_{sv}^\pi(t) = -\sum_{k=1}^{N} \left[ \sum_{\alpha,\beta=x,y,z} \kappa_k^{\alpha\beta} \hat{S}_\alpha \hat{S}_\beta \right] q_k(0) \exp\left( i \int_0^t w(r) dr \right), \tag{A6}$$

where $\left(w(r)\right)_{r=0}^{t} = \pi$. Assuming that no stochastic modulation occurs for any molecule in the entire ensemble between times $t$ and $t+\Delta t$, we have by equations (A3) to (A5),

$$f(\boldsymbol{\rho},t) = \left\langle f\left( \boldsymbol{\rho} - \frac{i}{\hbar} L^{\pi_k} \boldsymbol{\rho} \Delta t, t + \Delta t \big| \pi \right) \right\rangle. \tag{A7}$$

Expanding the right-hand side of equation (A7) to first order in $\Delta t$ gives

$$f(\boldsymbol{\rho},t) = \left\langle f(\boldsymbol{\rho},t|\pi) - \frac{i}{\hbar} \Delta t L^\pi \boldsymbol{\rho} \cdot \frac{\partial f(\boldsymbol{\rho},t|\pi)}{\partial \boldsymbol{\rho}} + \frac{\partial f(\boldsymbol{\rho},t|\pi)}{\partial t} \Delta t \right\rangle \tag{A8}$$

We now correct for stochastic modulation between times $t$ and $t+\Delta t$. For the molecules of the ensemble in subset $S$, consider a region $\Omega_t$ centered at point $\boldsymbol{\rho}$ with volume $\delta \boldsymbol{\rho}$ in the Hilbert space of the Liouville space vectors, and consider a second region $\Omega_{t+\Delta t}$ centered at point $\mathbf{\rho}-(i/\hbar)L^{\pi}\mathbf{\rho}\Delta t$ also with volume $\delta\mathbf{\rho}$. At time $t$, the Liouville space vectors of the molecules of $S$ are all contained in the region $\Omega_t$, and in the absence of stochastic modulation they will all be in region $\Omega_{t+\Delta t}$ at time $t+\Delta t$ as well. However, in the presence of stochastic modulation, some of the systems in $S_k$ will deviate from the point $\mathbf{\rho}-(i/\hbar)L^{\pi}\mathbf{\rho}\Delta t$, fall outside of the region $\Omega_{t+\Delta t}$ at time $t+\Delta t$, and the conservation equation (A4) will not hold. To account for the loss of probability density, we modify (A8) to

$$f(\mathbf{\rho},t) = \left\langle f(\mathbf{\rho},t|\pi) - \frac{i}{\hbar}\Delta t L^{\pi}\mathbf{\rho} \cdot \frac{\partial f(\mathbf{\rho},t|\pi)}{\partial \mathbf{\rho}} + \frac{\partial f(\mathbf{\rho},t|\pi)}{\partial t}\Delta t - Y(\mathbf{\rho},t,\Delta t|\pi) \right\rangle \quad (A9)$$

where $Y(\mathbf{\rho},t,\Delta t|\pi)$ is the loss of conditional probability density at point $\mathbf{\rho}-(i/\hbar)L^{\pi}\mathbf{\rho}\Delta t$ due to stochastic modulation on the systems in $S$ during the time period $t+\Delta t$. We obtain the stochastic Boltzmann equation as

$$\frac{\partial f(\mathbf{\rho},t)}{\partial t} - \frac{i}{\hbar}\left\langle L^{\pi}\mathbf{\rho} \cdot \frac{\partial f(\mathbf{\rho},t|\pi)}{\partial \mathbf{\rho}} \right\rangle = \left.\frac{\partial f(\mathbf{\rho},t)}{\partial t}\right|_{coll}, \quad (A10)$$

where the 'collision term' on the right-hand side is defined as

$$\left.\frac{\partial f(\mathbf{\rho},t)}{\partial t}\right|_{coll} = \lim_{\Delta t \to 0} \frac{\left\langle Y(\mathbf{\rho},t,\Delta t|\pi) \right\rangle}{\Delta t}. \quad (A11)$$

## ii. Appendix 2. Collision Term for Boson Gas Coupling

We now show that the boson gas coupling approach for simulating the stochastic frequency processes for the vibrational modes described in section 3 leads to a collision term in equation (A4) that satisfies the relaxation time approximation. As before, assume that the molecule only contains one vibrational mode. As mentioned in section 3, the boson gas coupling approach runs as follows. The vibrational mode frequency is of the form

$$W(t) = \hbar\omega(V(t) + 1/2), \tag{A12}$$

where $V(t)$ is a zero or positive integer-valued stochastic process. The process $V$ either increases by one unit ('stimulated absorption') or decreases by one unit ('stimulated emission') with equal probability at the end of random time intervals $T_1$, $T_2$, ...., subject to the constraint that $V(t) \geq 0$. For each $j$, $T_j$ follows an exponential distribution with rate parameter $\lambda$. The process $V$ also decreases by one unit ('spontaneous emission') at the end of random time intervals $U_1$, $U_2$, ... where for each $j$, $U_j$ is an exponential random variable with rate parameter $\gamma$. The sequences $U_1$, $U_2$, ... and $T_1$, $T_2$, ..., are independent.

Using the notion from part i above, fix a path $\pi$ for the frequency process $W$ and consider the systems $S$ as they travel from region $\Omega_t$ to region $\Omega_{t+\Delta t}$ between times $t$ and $t+\Delta t$. For sufficiently small $\Delta t$, the fraction of molecules in $S$ that do not land within $\Omega_{t+\Delta t}$ due to the stochastic modulation is simply the fraction of molecules that undergo a single stimulated absorption, stimulated emission, or spontaneous absorption event between times $t$ and $t+\Delta t$ (we assume that $\Delta t$ is small enough to that multiple transitions within the time period $\Delta t$ can be neglected). We therefore have

$$Y(\mathbf{\rho},t,\Delta t|\pi)\delta\mathbf{\rho} = \Pr\nolimits_{\pi}(U_1 < \Delta t \text{ or } T_1 < \Delta t) f(\mathbf{\rho},t|\pi)\delta\mathbf{\rho}, \qquad (A13)$$

where $\Pr_{\pi}(U_1 < \Delta t \text{ or } T_1 < \Delta t)$ is the probability that a stimulated emission or absorption process or a spontaneous emission process occurs within the time interval, and the subscript $\pi$ is to indicate that we are considering the path $\pi$. This probability can be decomposed into

$$\Pr\nolimits_{\pi}(U_1 < \Delta t \text{ or } T_1 < \Delta t) = \Pr\nolimits_{\pi}(U_1 < \Delta t) + \Pr\nolimits_{\pi}(T_1 < \Delta t) - \Pr\nolimits_{\pi}(U_1 < \Delta t \text{ and } T_1 < \Delta t).$$

(A14)

The third term on the right-hand side of (A14) can be ignored, because we have assumed that $\Delta t$ is too small for multiple transition events to occur with appreciable probability. According to the previous paragraph, we have $\Pr_{\pi}(U_1 < \Delta t) = 1 - \exp(-\lambda \Delta t)$ and $\Pr_{\pi}(T_1 < \Delta t) = 1 - \exp(-\gamma \Delta t)$. Substituting these into (A14) and expanding the exponentials to first order, we obtain

$$\Pr_\pi(U_1 < \Delta t \text{ or } T_1 < \Delta t) = \upsilon_\pi \Delta t, \quad (A15)$$

where

$$\upsilon_\pi = (2\lambda + \gamma). \quad (A16)$$

Substituting (A16) into (A13), and then substituting the result in (A11), gives an expression for the collision term

$$\left.\frac{\partial f(\mathbf{\rho},t)}{\partial t}\right|_{coll} = \langle \upsilon_\pi f(\mathbf{\rho},t|\pi) \rangle. \quad (A16)$$

Note that it is important to keep $\upsilon_\pi$ inside the brackets in (A16), as emission processes will not be available for frequency paths that have $V(t) = 0$ and so for these cases we have $\upsilon_\pi = \lambda$ in place of (A15).

Insight into the physical meaning of the relaxation time approximation can be acquired by considering what happens when multiple transition events can occur during the time interval $\Delta t$. For simplicity, suppose that exactly one or two transition events can occur with non-negligible probability during the interval $\Delta t$. We can replace equation (A13)

with

$$Y(\boldsymbol{\rho},t,\Delta t|\pi)\delta\boldsymbol{\rho} = f(\boldsymbol{\rho},t|\pi)\delta\boldsymbol{\rho}\left[q_1^{\pi}(\Delta t) + q_2^{\pi}(\Delta t)\right] \quad (A17)$$

where $q_n^{\pi}(\Delta t)$ is the probability that path $\pi$ will undergo exactly $n$ transition events over the time interval $\Delta t$. $q_1^{\pi}(\Delta t)$ is identical to formula (A15) derived above. As for $q_2^{\pi}(\Delta t)$, we have

$$q_2^{\pi}(\Delta t) = \Pr(U_1 + U_2 < \Delta t) + \Pr(T_1 + T_2 < \Delta t) + \Pr(\max(U_1, T_1) < \Delta t) \quad (A18)$$

By applying various rules concerning sums of exponential random variable, we can obtain (to lowest order in $\Delta t$),

$$\Pr(U_1 + U_2 < \Delta t) = \lambda^2 (\Delta t)^2, \quad (A19)$$

$$\Pr(T_1 + T_2 < \Delta t) = \gamma^2 (\Delta t)^2, \quad (A20)$$

and

$$\Pr(\max(U_1, T_1) < \Delta t) = \lambda\gamma(\Delta t)^2. \quad (A21)$$

Substituting these results (A19) – (A21) into (A17) then gives a collision term with

relaxation time constant

$$\upsilon_\pi = 2\lambda + \gamma + 4\lambda^2 (\Delta t)^2 + \gamma^2 (\Delta t)^2 + 2\lambda\gamma (\Delta t)^2 \tag{A22}$$

Equation (A22) demonstrates that the relaxation time approximation will break down when the frequencies $\lambda$ and $\gamma$ are large compared to the time scale $\Delta t$ over which the system evolves. This point is discussed further in Section 5 of the main paper.

**Ethics statement.** This research poses no ethical considerations.

**Data accessibility statement.** This work does not have any experimental data.

**Competing interests statement.** The authors have no competing interests.

**Authors' contributions.** DP and WT conceived the research. DP developed the theory. DP and HK performed the numerical simulations. DP, HK, and WT wrote the paper. All authors gave final approval for publication.

**Acknowledgements.** This research was supported by the World Premier Research Institute Initiative promoted by the Ministry of Education, Culture, Sports, Science, and Technology of Japan (MEXT) for the Advanced Institute for Materials Research, Tohoku University, Japan. We thank Dr. Kelly Reaves, Dr. Ikutaro Hamada, and participants of the program 'Materials for a Sustainable Energy Future' at IPAM, University of California, Los Angeles, for helpful discussions.

**Funding**. This research received no direct funding.